\documentclass[%
 reprint,
superscriptaddress,
preprintnumbers,
nofootinbib,
 amsmath,amssymb,
 aps,
]{revtex4-2}

\usepackage{graphicx}
\usepackage{dcolumn}
\usepackage{bm}
\usepackage{xcolor}

\newcommand{\cpp}[2][-]{\hat{c}_{#1 \bf #2}^{\dagger}} 
\newcommand{\cm}[2][]{\hat{c}_{#1 \bf #2}} 

\newcommand{\vt}[1]{{\small \bf #1}} 
\newcommand{\dpr}[2]{{\bf #1} \cdot {\bf #2}} 
\newcommand{\Mp}{M_{\rm Pl}} 

\begin{document}

\title{Quantum coherence of photons at cosmological distances}

\author{Arjun Berera}
 \affiliation{School of Physics and Astronomy, University of Edinburgh,
Edinburgh, EH9 3FD, United Kingdom}
\author{Suddhasattwa Brahma}%
\affiliation{%
 Department of Physics, McGill University, Montr\'eal, QC, H3A 2T8, Canada
}%
\author{Robert Brandenberger}
\affiliation{Department of Physics, McGill University, Montr\'eal, QC, H3A 2T8, Canada
}%
\author{Jaime Calder\'on-Figueroa}
\affiliation{School of Physics and Astronomy, University of Edinburgh,
Edinburgh, EH9 3FD, United Kingdom}
\author{Alan Heavens}
\affiliation{Imperial Centre for Inference and Cosmology (ICIC), Imperial College, Blackett Laboratory, Prince Consort Road, London, SW7 2AZ, United Kingdom}

\date{\today}

\begin{abstract}
We identify potential sources of decoherence for $U(1)$ gauge bosons from a cosmological standpoint. Besides interactions with different species in the cosmological medium, we also consider effects due to the expansion of the Universe, which can produce particles (especially scalars) than can potentially interact with the photon in a quantum state. We look in particular at the case of axion-like particles and their predicted decay channels in our analysis. These interactions are shown to have a negligible effect as far as decoherence goes. Interaction rates with CMB radiation or through Thomson scattering are small, so that the interstellar medium remains the biggest decoherence factor. Thus, quantum teleportation experiments with photon energies in the range $1$--$10$ keV should be feasible at cosmological distances up to the galaxy formation epoch or beyond ($z \sim 100$).
\end{abstract}

\maketitle


\section{Introduction}

Quantum teleportation experiments have shown that quantum coherence can be 
maintained for ever increasing distances. Indeed, the factor that hinders 
coherence (breaking the required entanglement for teleportation) is the 
loss of signal to the medium, mainly the atmosphere. This obstacle is no 
longer present in space, which hints at the possibility of performing 
such experiments at interstellar distances, or even detecting quantum 
signals from astrophysical sources. In this context, one of us showed 
recently that the quantum state of a photon could indeed be maintained 
at galactic distances, at least for a range of the electromagnetic 
spectrum \cite{Berera2020}. The reason for this is that the mean free 
paths associated to the different interactions the photon could have are 
many order of magnitude larger than the galactic scales 
(or even the observable Universe). 
As an outcome of this observation, one seminal suggestion that paper 
made was the possibility
for interstellar quantum communication, due to the viability for
maintaining quantum coherence over these distances
for certain frequency bands. Another possibility suggested in
that paper was if there were any natural quantum coherent sources,
such signals could maintain their coherence over interstellar
distances.   Extending on these ideas,
that paper also noted that this (lack of) effect most likely can be extrapolated to cosmological distances. 

This work will explore that possibility.  Here we consider
a wider variety of decoherence factors, like the expansion of the Universe itself. However, even for this case we do not give up on the philosophy that decoherence takes place due to the interaction of the quantum state with some environment. To do so, we consider the environment to be constituted by particles produced by the expansion of the Universe at different epochs. The mechanism to achieve this is squeezing, which has been widely studied in quantum optics and, in cosmology, in the theory of inflationary perturbations. So, borrowing from this mechanism, we compute the number of scalar particles through squeezing, and argue that this effect is essentially absent for fermions and $U(1)$ gauge bosons. Moreover, we identify the scalar field (interacting with photons) to be that of axion-like particles (ALPs), as a natural extension of the Standard Model. With these considerations, we are able to look at different interactions of the photon with the ALPs (or their decay products) in order to estimate the probability of interactions, which we find to be basically null. Thus, in practice, the expansion is not a decoherence factor for photons (at the energies we shall consider). We also look at other potential sources of decoherence, like interaction with CMB radiation or with electrons after reionization. The latter is more likely to be a source of decoherence, although the probabilities remain low enough to consider that the quantum state could remain undisturbed after decoupling. This opens up a new window to look for quantum signals from certain astrophysical objects or even from cosmic strings. 

\section{Expansion-induced decoherence}

\subsection{Scalar fields}

In order to learn how the expansion of the Universe can lead to decoherence, let us look at the theory of cosmic inflation for guidance. 
Cosmological perturbations during inflation undergo a process known as {\it squeezing}, where states of the type $|n_{\bf k}, n_{-{\bf k}}\rangle$ are created at superhorizon scales. This is an effect purely due to expansion, whose basic principles can be grasped just by studying a massless scalar field minimally coupled to gravity, as follows:
\begin{align}
	{\cal S} & = \frac{1}{2} \int dt\ d^3 x\ \sqrt{-g} \partial_{\mu} \phi \partial^{\mu} \phi \nonumber \\
	& = \frac{1}{2} \int d\tau\ d^3 x\ a^2 \left[(\phi')^2 - (\nabla \phi)^2 \right],
\end{align}
where primes denote derivative w.r.t. the conformal time $\tau$. It is convenient to introduce the change of variable $\varphi \equiv a \phi$, such that
\begin{equation}\label{act1}
	{\cal S} = \frac{1}{2}\int d\tau\ d^3 x\ \left[ \left(\varphi' - \frac{a'}{a}\varphi\right)^2 - (\nabla \varphi)^2 \right].
\end{equation}
Using the Euler-Lagrange equations, and going to Fourier space, one gets the mode equations
\begin{equation}\label{eom1}
	\varphi_k'' + \left(k^2 - \frac{a''}{a}\right) \varphi_k = 0.
\end{equation} 
In the case of a perfect de Sitter expansion, $a''/a = 2/\tau^2$, the equation of motion becomes
\begin{equation}
	\varphi_k'' + \left(k^2 - \frac{2}{\tau^2} \right) \varphi_k = \varphi_k'' + \left(k^2 - 2 (aH)^2 \right) \varphi_k = 0\,.
\end{equation}
Clearly, the solutions are oscillatory for $k^2 > 2 (aH)^2$, whereas for $k^2 < 2(aH)^2$ there is a growing and a decaying-mode solution. The question which then arises is what should be the right initial state for solving this equation. For inflation, one usually takes Bunch-Davies initial states
\begin{equation}
	\varphi_k (\tau) = \frac{e^{-i k \tau}}{\sqrt{2k}} \left(1 - \frac{i}{k\tau}\right)\,,
\end{equation}
such that the time-dependent field operator and the canonical momentum are given by
\begin{widetext}
\begin{equation}
	\hat{\varphi} (\tau,\vt{x}) = \int \frac{d^3k}{(2\pi)^3} \frac{1}{\sqrt{2k}} \left[e^{-ik\tau}\left(1 - \frac{i}{k\tau}\right) \cm{k} (\tau_0) + e^{ik\tau} \left(1 + \frac{i}{k\tau}\right) \cpp{k} (\tau_0) \right] e^{i \dpr{k}{x}},
\end{equation}
\begin{equation}
\hat{\pi}(\tau, \vt{x}) = \varphi' - \frac{a'}{a}\varphi = -i \int \frac{d^3k}{(2\pi)^3} \sqrt{\frac{k}{2}} \left[e^{-ik\tau} \cm{k} (\tau_0) - e^{ik\tau} \cpp{k} (\tau_0)\right] e^{i \dpr{k}{x}}\,.
\end{equation}
\end{widetext}

The creation and annihilation operators at later times can be found through a Bogolyubov transformation, such that
\begin{align}
	\cm{k} (\tau) & = \alpha_k (\tau) \cm{k} (\tau_0) + \beta_k (\tau) \cpp{k}(\tau_0)\,, \nonumber\\
	\cpp{k} (\tau) & =  \alpha_k^* (\tau) \cpp{k} (\tau_0) + \beta_k^* (\tau) \cm{k} (\tau_0)\,,
\end{align}
where $|\alpha_k|^2 - |\beta_k|^2 = 1$.

Considering this, one can parametrize these coefficients as
\begin{equation}
	\alpha_k = \cosh (r_k) e^{-i \Theta_k},  \quad \beta_k = -\sinh (r_k) e^{i (\Theta_k + 2\phi_k)}\,,
\end{equation}
which renders
\begin{widetext}
\begin{align}\label{sqpar}
	\hat{\varphi}_k (\tau) =& \frac{1}{\sqrt{2k}} \left\{\left[ \cosh (r_k) e^{-i \Theta_k} - \sinh(r_k) e^{-i(\Theta_k + 2\phi_k)}\right] \cm{k} + \left[\cosh(r_k) e^{i\Theta_k} - \sinh(r_k) e^{i(\Theta_k + 2\phi_k)}\right] \cpp{k} \right\}, \nonumber \\
	\hat{\pi}_k (\tau) = & -i \sqrt{\frac{k}{2}} \left\{\left[ \cosh (r_k) e^{-i \Theta_k} + \sinh(r_k) e^{-i(\Theta_k + 2\phi_k)}\right] \cm{k} - \left[\cosh(r_k) e^{i\Theta_k} + \sinh(r_k) e^{i(\Theta_k + 2\phi_k)}\right] \cpp{k} \right\}\,.
\end{align}
\end{widetext}

Comparing with the equations above (depending on Bunch-Davies functions), one readily finds the parameters
\begin{align}
	r_k & = \sinh^{-1} \left(\frac{1}{2k\tau}\right), \qquad \Theta_k = k\tau + \tan^{-1}  \left(\frac{1}{2k\tau}\right)\,, \nonumber \\
	\phi_k & = -\frac{\pi}{4} - \frac{1}{2} \tan^{-1}  \left(\frac{1}{2k\tau}\right).
\end{align}
The vacuum expectation value of the number of particles for the new vacuum in the $k$ mode  is given by
\begin{equation}
	\left\langle N_k \right\rangle = |\beta_k|^2 = \sinh^2 (r_k) = \left(\frac{1}{2k\tau}\right)^2.
\end{equation}
Thus, for $k < 2/(aH)$ the expectation number is bigger than $1$. In practice, this matches the region for which the equation of motion has the exponential solutions, and in particular, where squeezing takes place. 

\subsubsection{Particle production during the standard cosmological expansion}

In order to find the density of particles created during the expansion history, it is convenient to have at hand the evolution of the scale factor as a function of the conformal time, starting from the inflationary era until the matter dominated era. We shall assume the transitions between epochs to be instantaneous, commonly known as the sudden approximation.\footnote{Relaxing this assumption does not change our main findings.} Using the sudden approximation between the (quasi)-de Sitter expansion and the hot big bang phase, the scale factor is given by

\begin{equation*}
	a(\tau)=\left\{\begin{array}{l}
(H_{\rm Inf}|\tau|)^{-1}, \quad \tau<\tau_{e}<0 \\
\alpha_M (\tau-\tau_e)^2 + \alpha_R (\tau - \tau_e) + \alpha_I, \quad \tau>\tau_{e}
\end{array}\right.\,,
\end{equation*}
\begin{align}
	\alpha_M & = \frac{\pi G}{3} \rho_{eq} a_{eq}^3, \quad \alpha_I = \frac{1}{H_{\rm Inf} |\tau_e|}, \nonumber \\
	 \alpha_R & = \left[ \frac{4\pi G}{3} \rho_{eq} a_{eq}^3 \left(\frac{1}{H_{\rm Inf} |\tau_e|} + a_{eq} \right)\right]^{1/2}\,,
\end{align}
where  $\tau_e$ denotes the conformal time at the end of inflation and ``$eq$'' refers to the time of matter-radiation equality. The quadratic term corresponds to the evolution during matter domination, whereas the linear term to radiation domination. 

Then, the equation of motion Eq. \eqref{eom1} (which is for a completely general cosmological background) during this epoch(s) is given by
\begin{equation}
	\varphi_k'' + \left(k^2 - \frac{2\alpha_M}{\alpha_M(\tau-\tau_e)^2 + \alpha_R(\tau-\tau_e) + \alpha_I} \right) \varphi_k = 0\,.
\end{equation}
Naturally, one can identify regions where the equation gets simplified. For the radiation-dominated era, the e.o.m. is
\begin{equation}
	\varphi_k'' + k^2 \varphi_k = 0\,,
\end{equation}
whereas for the matter-dominated era, it is given by
\begin{equation}
	\varphi_k'' + \left(k^2 - \frac{2}{\tau^2} \right) \varphi_k = 0\,,
\end{equation}
i.e., the same equation as for the inflationary era. In principle, there can be small changes to the usual (positive-frequency) vacuum state coming from effects of gravitational phase transitions. However, the corrections to the positive-frequency vacuum are small or, in other words, the number of particles created due to these phase transitions quickly dilute. Therefore, it is reasonable to consider the Bunch Davies-like initial states, such that the solutions to these equations are, respectively, given by
\begin{eqnarray}
{}^{\rm Rad}\varphi_k (\tau) = \frac{1}{\sqrt{2k}} e^{-ik/\mathcal{H}}\,,
\end{eqnarray}
and the basis for the matter-dominated era as
\begin{eqnarray}\label{solm}
	{}^{\rm Mat}\varphi_k (\tau) = \frac{1}{\sqrt{2k}} \left(1 - \frac{i \mathcal{H}}{2k}\right) e^{-2ik/\mathcal{H}}\,,
\end{eqnarray}
where ${\cal H}$ is the comoving rate of expansion.\footnote{We here ignore any
    squeezing which takes place before the epoch of radiation domination.} As mentioned, the matching of the solutions during the different epochs will lead to excited states that will increase the number of generated particles. However, for now it will be enough to concentrate on this simple form of solutions. Moreover, notice that Eq. \eqref{solm} has the same functional form as the Bunch-Davies solution for de Sitter spacetime, and thus the squeezing formalism derived for that case also applies here. In particular, the vacuum expectation of the number of particles is 
\begin{equation}
	\langle N_k \rangle = |\beta_k|^2 = \left(\frac{1}{2k\tau}\right)^2\,.
\end{equation}
On the contrary, during radiation domination there is no mass term in the e.o.m., so there is no squeezing and particle production during this era. Therefore, expansion induces particle excitations of a scalar field only during the de Sitter and matter-dominated eras. Before moving on, we will cover the case of massive scalar fields during inflation, and similar calculations can be done for standard expansion.

\subsubsection{The massive scalar case}

Here we will cover (although somewhat superficially) the case of a massive scalar field. A priori, one would expect particle production for massive fields to be less efficient, so we need to quantify the required corrections to the functions displayed above.

Let us start with a generalization of the action Eq. \eqref{act1},
\begin{equation}
	{\cal S} = \frac{1}{2} \int d\tau d^3 x \left[ \left(\varphi' - \frac{a'}{a}\varphi \right)^2 - (\nabla \varphi)^2 - a^2 m^2 \varphi^2 \right]\,,
\end{equation}
which renders the following equation of motion for the field
\begin{equation}
	\varphi_k'' + \left[k^2 - \left( \frac{a''}{a} - a^2 m^2 \right) \right] \varphi_k = 0\,.
\end{equation}
Once again, this equation is completely general for any cosmological epoch. For inflation this becomes
\begin{equation}
	\varphi_k'' + \left[k^2 - \frac{1}{\tau^2} \left(2 - \frac{m^2}{H^2}\right) \right]\varphi_k = 0\,,
\end{equation}
where the Bunch-Davies solution is
\begin{equation}\label{fim}
	\varphi_k = \frac{e^{i (2\nu+1)\frac{\pi}{4}}}{\sqrt{2k}} \sqrt{\frac{\pi}{2}} w^{1/2} H_{\nu}^{(1)} (w)\,,
\end{equation}
where $w = |k \tau|$ and $\nu^2 = 9/4 - (m/H)^2$. The conjugate momentum is
\begin{align}\label{pim}
	\pi_k = & -i \sqrt{\frac{k}{2}} \bigg\{-i \sqrt{\frac{\pi}{2}} e^{i (2\nu+1)\frac{\pi}{4}} \bigg[ w^{1/2} H_{\nu-1}^{(1)} (w) \nonumber \\
	& + w^{-1/2} \left(3/2 - \nu \right) H_{\nu}^{(1)} (w) \bigg] \bigg\}\,.
\end{align}
Notice that negative values of $\nu$ lead to exponentially suppressed solutions. Thus, as expected, there is no particle production for $m \gtrsim H$. Then, assuming that the mass term is small enough so that $\nu$ is safely larger than 0, one can compute the number of generated particles due to squeezing by comparing the equations above with Eq. \eqref{sqpar}. In order to have analytical expressions, one can expand eqs.\eqref{fim},\eqref{pim} in powers of $m/H$, which yields
\begin{align}
	|\beta_k|^2 \simeq \left(\frac{1}{2k\tau}\right)^2 \left[1 + \frac{2}{3} \frac{m^2}{H^2} \left( -1 + \gamma_E + \ln (2w) \right)\right]\,,
\end{align}
where $\gamma_E$ denotes the Euler-Mascheroni constant. Naturally, during radiation domination this type of mass term does not enhance squeezing, whereas during matter dominance it is more subdominant than in the other eras ($\tau^{-4}$ vs. $\tau^{-2}$). 

\subsection{Setting up an environment}

What we have covered so far is valid for a scalar field, so the natural next step is to try and reproduce this for photons. However, in this case there is no induced time-dependent mass-term and thus no squeezing (similarly to the scalar case during radiation domination). Naturally, there can be particle production due to interactions with other fields, but such processes are not linked to the background dynamics. In fact, in some cases the expansion just dilutes whatever number of particles are produced through these couplings. Consequently, in order to grasp the effects of decoherence of photons due to expansion alone, the next best thing is to look at the interactions between the quantum state (of a test photon) and an environment encompassed by either pseudoscalar particles produced by the squeezing of super-horizon states, or by decay products of these scalars, in particular, into photons. Arguably, the preeminent example of a scalar field in such scenario is the axion, which has a well-known interaction with $U(1)$ fields. Moreover, the interactions between axions and photons through other means have been widely explored in the literature, where the search of this particle is largely based on this interaction. 
The interaction between axions and $U(1)$ gauge fields is described by the Lagrangian
\begin{equation}
	{\cal L}_{A\gamma\gamma} = - \frac{g_{A \gamma\gamma}}{4} F_{\mu\nu} \tilde{F}^{\mu\nu} \phi_A = g_{A\gamma\gamma} \vt{E}\cdot \vt{B}\ \phi_A\,,
\end{equation}
with
\begin{align}
	g_{A\gamma\gamma} & = \frac{\alpha}{2\pi f_A} \left(\frac{E}{N} - 1.92(4)\right) \nonumber \\
	& = \left(0.203(3) \frac{E}{N} - 0.39(1) \right) \frac{m_A}{{\rm GeV}^2}\,,
\end{align}
where $E$ and $N$ are the electromagnetic and color anomalies of the axial current \cite{RRR2018}. 

\subsubsection{Number density}

Let us estimate the number density of $\phi_A$-particles created during inflation (just by squeezing). For this, we need to compute the total number of particles. Assuming the states are homogeneously distributed, the amount of states within a ``radius'' $k$ is 
\begin{equation}
	G(k) = \frac{V}{(2\pi)^3} \frac{4\pi k^3}{3}\,,
\end{equation}
where $V$ stands for a comoving volume. In this way, the density of states is given by
\begin{equation}
	g(k) = \frac{\partial G}{\partial k} = \frac{V}{2\pi^2} k^2\,.
\end{equation}
Thus, the total number of particles is
\begin{widetext}
\begin{equation}
	N = \sum_k N_k = \int dk\ g(k) f(k) \nonumber  = \frac{V}{2\pi^2} \int_{0}^{-1/\tau_e}  dk\ k^2 \left(\frac{1}{2k\tau_e}\right)^2 \left[1 + \frac{2}{3} \frac{m^2}{H_{\rm Inf}^2} \left( -1 + \gamma_E + \ln (-2k\tau_e) \right)\right]\,,
\end{equation}
\end{widetext}
where we have used the formula for the average number of particles on a mode $k$ created due to squeezing, which we identified with $f(k)$. The integration limits correspond only to modes that have been superhorizon at some point during inflation, as those are the ones that undergo squeezing. Performing the integral we get
\begin{eqnarray}
	N & = & \frac{V}{8\pi^2} \left. \frac{k}{\tau_e^2} \left[1 + \frac{2}{3}\frac{m_a^2}{H_{\rm Inf}^2} \left( -2 + \gamma_E + \ln (-2k\tau_e)\right] \right]\right|_{0}^{-1/\tau_e} \nonumber \\
	& = & -\frac{V}{8\pi^2} \frac{1}{\tau_e^3} \left[1 + \frac{2}{3}\frac{m_a^2}{H_{\rm Inf}^2} \left( -2 + \gamma_E + \ln 2 \right)  \right]\,.
\end{eqnarray}
Then, one can obtain the density of these particles at any given time through
\begin{equation*}
	n = \frac{N}{V_{\rm phys}} = - \frac{1}{8\pi^2} \left(\frac{1}{a \tau_e}\right)^3 \left[1 + \frac{2}{3} \frac{m_a^2}{H_{\rm Inf}^2} \left( -2 + \gamma_E + \ln 2 \right) \right]\,.
\end{equation*}

This is a good point to make some estimates. First, one can get away (for now) with not choosing a value of $m_a$, as it will be subdominant. Thus, we are left to find $\tau_e$. To do so, notice that 
\begin{equation}
	\frac{1}{k_0 |\tau_e|} = \frac{k_0|\tau_*|}{k_0 |\tau_e|} = \frac{a_e}{a_*} \sim e^{60}\,,
\end{equation}
where `$0$' and `$*$' stand for present-day and horizon-crossing magnitudes. In particular, $k_0$ can be identified with the current horizon length. As it is widely known, inflation had to last at least 60 $e-$folds after this mode crossed the horizon in order to solve the horizon problem.\footnote{Actually, the number of $e-$folds needed to solve the horizon problem depends on the energy scale of inflation, but we are only interested in rough estimates here.} Then, the above is equivalent to 
\begin{equation}
	\frac{(aH)_0^{-1}}{|\tau_e|} = \frac{H_0^{-1}}{|\tau_e|} \sim e^{60}\,,
\end{equation}
rendering a conformal time at the end of inflation,
\begin{equation}
	\tau_e \sim - 4 \times 10^{-9}\ {\rm sec} = - 1.465\times 10^{34} \Mp^{-1}\,.
\end{equation}
With this, we have the necessary values to estimate the number density of squeezing-generated ALPs at any given era. The free parameters are the energy scale of inflation and the mass of the particles. However, if the latter is small in comparison to the former, the contribution from the ratio will be negligible and one can get away with working with the first term. 

\subsection{$\phi_A \gamma_t \rightarrow x \overline{x}$}

Fermion production from the interaction of an ALP and a (test) photon $\gamma_t$ is mediated by the Lagrangian
\begin{equation}
	q_x A_{\mu} \overline{\psi} \gamma^{\mu} \psi\,.
\end{equation}

Let us take the initial momenta of the particles to be
\begin{equation}
	k_a = E_a(1, \cos \theta, \sin \theta ,0), \qquad k_{\gamma} = E_{\gamma}(1, 1, 0, 0)\,.
\end{equation}
With a center of mass energy given by $E^2_{\rm com} = 2 E_a E_{\gamma} (1-\cos \theta)$, one can find the cross section of the interaction to be
\begin{equation}
	\sigma = \frac{1}{4E_a E_{\gamma} |v_a - v_{\gamma}|} \frac{q_x^2 m_x^2}{2\pi f_a^2} \ln \left(\frac{E' + p'}{E'-p'}\right)\,,
\end{equation}
where $E' = E_{\rm com}/2$, $p' = \sqrt{(E')^2 - m_x^2}$ and $m_x$ denotes the mass of the fermions. Now, we introduce the variables $\lambda = 2 m_x^2/(E_a E_{\gamma})$ and $y = \cos \theta$, such that the average over the initial axion momentum is \cite{Conlon:2013isa}
\begin{widetext}
\begin{align}
	\langle \sigma v \rangle & = \frac{q_x^2 \lambda}{16\pi f_a^2} \int_{-1}^{1-\lambda} dy\ \ln \left(\frac{\sqrt{1-y} + \sqrt{1-y-\lambda}}{\sqrt{1-y} - \sqrt{1-y-\lambda}} \right) \nonumber \\
	& = \frac{q_x^2 \lambda}{16\pi f_a^2} \left[ - \sqrt{4-2\lambda} + (\lambda - 2) \ln (\sqrt{2} - \sqrt{2-\lambda}) + 2 \ln (\sqrt{2-\lambda} + \sqrt{2}) - \frac{1}{2} \lambda \ln \lambda \right] \nonumber \\
	& = \frac{q_x^2 \lambda}{16\pi f_a^2} \left[ - \sqrt{4-2\lambda} + (4-\lambda) \ln (\sqrt{2} + \sqrt{2-\lambda}) + \left(\frac{\lambda}{2} - 2\right) \ln \lambda \right]\,.
\end{align}
\end{widetext}
Notice this expression tells us that $0 < \lambda \leq 2$. 

Next, we identify two contributions to the ALP number density during the matter dominated era: those produced during inflation and those produced during matter domination itself. 
\begin{equation*}
	n = \int dn = \frac{1}{8\pi^2} \left[ \int_{aH}^{-1/\tau_e} \frac{dk}{a^3 \tau_e^2} + \int_{aH}^{(aH)_{eq}} \frac{dk}{a^3 \tau^2} \right]\,.
\end{equation*}
Then, it is convenient to write every expression in terms of the variable $\lambda$ introduced above, 
\begin{equation}
k = \frac{2 a m_x^2}{\lambda E_{\gamma}} \implies d k= - \frac{2am_x^2}{E_{\gamma}} \frac{d\lambda}{\lambda^2}\,,
\end{equation}
such that the interaction rate is given by
\begin{widetext}
\begin{align}
	\langle n \sigma v \rangle = & \frac{q_x^2}{16\pi f_a^2} \frac{1}{8\pi^2 a^3} \frac{2am_x^2}{E_{\gamma}} \bigg\{ \int_{\lambda_e}^{\lambda_{\tau}} \frac{d\lambda}{\lambda \tau_e^2} \left[ - \sqrt{4-2\lambda} + (4-\lambda) \ln (\sqrt{2} + \sqrt{2-\lambda}) + \left(\frac{\lambda}{2} - 2\right) \ln \lambda \right] \nonumber \\
	& + \int_{\lambda_{\tau}}^{\lambda_{eq}} \frac{d\lambda}{\lambda \tau^2} \left[ - \sqrt{4-2\lambda} + (4-\lambda) \ln (\sqrt{2} + \sqrt{2-\lambda}) + \left(\frac{\lambda}{2} - 2\right) \ln \lambda \right] \bigg\}\,,
\end{align}
\end{widetext}
where 
\begin{equation*}
	\lambda_e = \frac{2m_x^2}{E_{\gamma}} a|\tau_e|\,, \qquad \lambda_{\tau} = \frac{2m_x^2}{H E_{\gamma}}\,, \qquad \lambda_{eq} = \frac{2m_x^2}{H_{eq} E_{\gamma}}\,.
\end{equation*}
The kinematic constraints on $\lambda$ place stringent bounds on the allowed values of the parameters of the model, in particular on the ratio $m_x^2/E_{\gamma}$. To see this, take the values at matter-radiation equality, where
\begin{equation*}
	\lambda_e(a_{eq}) \sim 10^{30} \Mp^{-1} \frac{2m_x^2}{E_{\gamma}}\,, \quad \lambda_{\tau} (\lambda_{eq}) \sim 10^{55} \Mp^{-1} \frac{2m_x^2}{E_{\gamma}}\,.
\end{equation*}
So, taking the maximum allowed value of $\lambda$, we conclude that $m_x^2/E_{\gamma} \sim 10^{-55} \Mp$. This could be satisfied only for extremely light fermions (even for not-so-realistic values of the photon energy). Assuming these rather implausible conditions are satisfied, we can notice that the first integral will dominate ($\tau_e^{-2} \gg \tau^{-2}$), so we will just focus on this one (for now). Then, the interaction rate is
\begin{equation}
	\langle n \sigma v \rangle \approx \frac{3356 q_x^2}{16\pi f_a^2} \frac{(1+z_{eq})^2}{8\pi^2} \frac{2\times 10^{-55}\Mp}{\tau_e^2}\,, 
\end{equation}  
where we have solved the integral numerically. Plugging the numerical values of $\tau_e$ and $z_{eq}$, we have that 
\begin{equation}
	\langle n \sigma v \rangle \sim 10^{-112} \Mp^3 \frac{q_x^2}{128\pi^2 f_a^2}\,.
\end{equation}
Clearly $f_a$ would need to be abnormally small in order to have a non negligible interaction rate. The only way to obtain non negligible values would be to suppress even more the ratio $m_x^2/E_{\gamma}$, such that the corresponding versions of $\lambda$ approach to $0$, where the integral actually diverges. Needless to say, even considering very light fermions, the energy of the photon would be out of reach (and can even become trans-Planckian). Indeed, for axions coming from string theory, we generically expect $f_a>\Mp$ from the Weak Gravity Conjecture (WGC) \cite{Heidenreich:2015nta, Rudelius:2014wla, Bachlechner:2015qja}. Interestingly, the WGC also constrains the ratio of the charge-to-mass of fermions to be less than $q_x/m_x < 1$ in Planck units. On excluding trans-Planckian photons on physical grounds, this means that $q_x$ gets naturally suppressed on considering very small values for $m_x^2/E_{\gamma}$. Therefore, it seems that the WGC highly disfavours having a non-negligible value for this interaction rate.

\subsection{$\phi_A \rightarrow \gamma \gamma \implies \gamma_t \gamma \rightarrow \gamma \gamma$}

In this case, we will check how likely it is for the photon to interact with an environment composed of photons which are produced from the decay of an ALP. For this, we need the decay width of the process, which is 
\begin{equation}
	\Gamma_{A\rightarrow \gamma\gamma} = \frac{g_{A\gamma\gamma}^2 m_A^3}{64\pi}\,,
\end{equation}
and, assuming $E/N = 0$, this becomes
\begin{equation}
	\Gamma_{A\rightarrow \gamma\gamma} =  1.1 \times 10^{-24}\ {\rm s}^{-1} \left(\frac{m_A}{\rm eV}\right)^5\,.
\end{equation}
Without any further calculations, one can see that for masses $m_A \sim {\cal O}(1)\ {\rm eV}$ or less, the decay width is too small even considering the age of the Universe ($\sim 10^{17}\ {\rm s}$), and so no photons would be produced. Current bounds on the mass of the axion highly disfavour higher masses. This is why it is more appropriate to talk about ALPs, as they are more generic and well suited to be a test lab. 

Naturally, the photons resulting from the decaying of the ALP will not have the same momentum as it. We label the resulting photons as $1'$ and $2'$, with an angle $\theta'$ between their momenta. Then, one can easily show that 
\begin{equation}
	\langle \cos \theta' \rangle = - \frac{m_A^2}{4 p_{1'} p_{2'}}\,, \qquad \langle p_{1'} p_{2'} \rangle = \frac{m_A^2}{4}\,,
\end{equation}
so that
\begin{equation}
	p_{1'}^2 + p_{2'}^2 = p_A^2 + \frac{m_A^2}{2}\,,
\end{equation}
leading to the following direction-averaged momenta
\begin{align}
	p_{1'}^2 &= \frac{m_A^2}{4} + \frac{p_A^2}{2} \left[ 1 + \sqrt{1+\frac{m_A^2}{p_A^2}} \right]\,, \nonumber \\
	p_{2'}^2 &= \frac{m_A^2}{4} + \frac{p_A^2}{2} \left[ 1 - \sqrt{1+\frac{m_A^2}{p_A^2}} \right]\,.
\end{align}
This leads to a not-so-simple distribution of photons. However, considering the range of masses that render a photon population at matter domination, the distribution can be somewhat simplified. To see this, first notice that the comoving momentum is between $(aH)_{eq} \lesssim k \leq (aH)_{e}$, or plugging in numbers, $10^{-59}\ \Mp \lesssim k \lesssim 10^{-34}\ \Mp$. The physical momentum of massive particles varies with expansion the same way as for massless particles ($p \propto a^{-1}$). Thus, the physical momentum of ALPs should be on the range $10^{-56}\ \Mp \lesssim p \lesssim 10^{-31}\ \Mp$ (or $ 10^{-38}\ {\rm GeV} \lesssim p \lesssim 10^{-13}\ {\rm GeV}$). Even for the upper limit, the physical momentum of ALPs is rather negligible in comparison with the rest mass required for it to decay by the matter dominated era (${\cal O}(10^3)\ {\rm eV}$). Thus, it is a good approximation to treat the ALPs as non-relativistic. Then, the momentum of the resulting photons are roughly
\begin{equation}
	p_{1'} \approx \frac{m_A + p_A}{2}\,, \qquad p_{2'} \approx \frac{m_A - p_A}{2}\,,
\end{equation}
where for the sake of simplicity, we take $p_{1'} \approx p_{2'} \approx m_A/2$.

With these considerations, one can compute the mean free path of a test photon interacting with an environment of photons decaying from ALPs. For starters, Euler and Kockel computed the cross section for photon-photon interactions \cite{Euler:1935zz, Liang:2011sj}, 
\begin{equation}
	\sigma(\gamma\gamma \rightarrow \gamma\gamma) = \frac{937 \alpha^4 \omega^6}{10125 \pi m^8},
\end{equation} 
where $\alpha \simeq 1/137$ is the fine structure constant, $\omega$ is the energy of the photons in the center-of-momentum frame, and $m$ is the mass of the electron. The momentum of each photon in the lab-frame can be written as
\begin{equation}
p_1^{\mu} = E_{1} (1,1,0,0), \qquad p_2^{\mu} = E_2 (1, -\cos \theta, \sin \theta, 0),
\end{equation}
such that
\begin{equation}
	\omega = \sqrt{E_1 E_2}\; \cos \frac{\theta}{2} .
\end{equation}
Next, recalling the number density of ALPs (which translates into the number density of photons up to a factor of $2$), and considering that their mass is negligible in comparison to the energy scale of inflation, we have
\begin{equation}\label{npeq}
	n = - \frac{1}{4\pi^2} \left(\frac{1}{a \tau_e}\right)^3\,,
\end{equation}
such that
\begin{equation}
	\sigma n \sim \frac{937 \alpha^4}{10125\pi m^8} E_{\gamma}^3 E_{1}^3 \frac{2}{\pi} \frac{(1+z_{eq})^3}{4\pi^2} |\tau_e|^{-3}\,,
\end{equation}
where $E_{\gamma}$ denotes the energy of the test photon (quantum state) and $E_{1}$ the energy of the environment photon. Then, taking $E_{\gamma} = 10^{-17}\ \Mp$ and $E_{1} \sim m_A = 10^{-24}\ \Mp\ (1\ {\rm keV})$, the resulting mean free path is
\begin{equation}
	\ell = (\sigma n)^{-1} 	\sim 10^{21}\ {\rm cm}\,,
\end{equation}
which should be compared to $H_{eq}^{-1} \sim 10^{50}\ {\rm cm}$. Nevertheless, notice that we have taken a rather high energy for the test  photon, so much so that the cross section formula may be invalid due to other processes being predominant. A more sensible value would be $E_{\gamma} = 10^{-24}\ \Mp$, which yields
\begin{equation}
	\ell = (\sigma n)^{-1} 	\sim 10^{42}\ {\rm cm}\,.
\end{equation}
Thus, in principle photons could interact with other photons emerging from the decay of ALPs (we will check this more carefully below). However, it is instructive to compare the possibility of these interactions to the interaction with CMB photons. According to our estimation for the number density of photons created through the process $\phi_A \rightarrow \gamma\gamma$, by the time of photon decoupling we have $ n \sim 20 \ {\rm cm}^{-3}$ $(600\ {\rm cm}^{-3}$ by matter-radiation equality), whereas for CMB photons $n_{pd} \approx n_{\gamma,0} (1+z_{pd})^3 \sim 4 \times 10^{11}\ {\rm cm}^{-3}$. Thus, the number density of ALP photons is negligible in comparison to CMB photons, so the latter are in principle a more important source of decoherence than the former after $z \sim 1000$. Let us compute next the mean free path due to this interaction. 

\subsection*{Mean free path}

In order to compute the mean free path (or redshift in a cosmological setting), we will use the optical depth, defined as 
\begin{equation}
	{\cal T} = \int \sigma j_{\mu} dx^{\mu}\,,
\end{equation}
where $\sigma$ is the cross section of the interaction and $j_{\mu}$ is the four-current \cite{Ruffini:2015oha}. The integral over the spatial dimensions are null due to isotropy and homogeneity. 
This will be used to compute in a more robust manner the mean free path for the interaction of a photon with others produced by the decay of an ALP. Moreover, we will incorporate the time dependence from the decay width. With these considerations, the optical depth is written as 
\begin{widetext}
\begin{equation}\label{opd}
	{\cal T} = \int_{t}^{t_0} dt\ (1-e^{-\Gamma t}) \frac{937\alpha^4 E_{\gamma}^3 m_A^3}{10125\pi m^8} \int_{-1}^{1} d(\cos \theta) \cos^6 \frac{\theta}{2}\ \frac{1}{4\pi^2} \left[\int_{aH}^{-1/\tau_e} \frac{dk}{\tau_e^2} + \int_{aH}^{(aH)_{eq}} \frac{dk}{\tau^2} \right]\,.
\end{equation}
\end{widetext}

Next, we shall assume a matter dominated Universe throughout the entire propagation of the photon. This will be convenient in order to deal with the explicit time dependence in the expression. Thus, we have that
\begin{equation}
	t = \frac{2}{3H_0} (1+z)^{-3/2}\,.
\end{equation}

We shall focus on the first term inside the brackets of \eqref{opd}, which is dominant (by many orders of magnitude). In doing so, the optical depth is written as
\begin{align}
{\cal T} & = \frac{937\alpha^4 E_{\gamma,0}^3 m_A^3}{10125\pi m^8} \frac{1}{8\pi^2} \times \nonumber \\ & \int_0^z \frac{dz'\ (1+z')^3}{H_0 (1+z')^{5/2}} (1-e^{-\Gamma t})\frac{1}{\tau_e^2}\left[-\frac{1}{\tau_e} - H_0 (1+z)^{1/2}\right]\,. \nonumber
\end{align}
In order to have numerical estimates we take $E_{\gamma,0} = m_A = 10^{-24}\ \Mp$, such that
\begin{align}
	\frac{937\alpha^4 E_{\gamma,0}^3 m_A^3}{10125\pi m^8} \frac{1}{8\pi^2 H_0} \simeq 10^{78}\,.
\end{align}
The probability of the photon travelling without interacting with the environment is given by $P(z) = e^{- {\cal T}(z)}$. For $z = 3400$, one gets ${\cal T} \sim 10^{-20}$, meaning that basically $P = 1$, and so there is no decoherence due to the interaction between the photon in some quantum state and the photons produced by the decay of expansion-generated ALPs. One could entertain the idea of going further into the past (higher redshift) in order to obtain non-trivial probabilities (even though the single-fluid approximation would break in the realistic setup). However, even for redshifts as high as $10^{20}$, the optical depth is just around $10^{-13}$, so that interactions remain highly unlikely. One could also argue that different input parameters could change this conclusion, however, smaller masses only lead to less efficient interactions and a slower decay, effectively increasing the mean free path. 

Let us emphasize that we have studied the potential interactions with particles that have been produced directly or indirectly due to the dynamics of the expansion of the Universe. In this sense, one could also ask if there can be interactions with a primordial population of ALPs (or their offspring). Such interactions can be potentially more important that the ones we have considered; however, it has been found that for realistic values of the parameters the growth of the photon field in particular is strongly suppressed \cite{Garretson1992, Arza2020}, and thus by the time of decoupling this scenario should not be considered a source of decoherence. 

An interesting thing to note is that the strength of the interaction, which we have considered in this work, has recently been constrained from the observation of the birefringence angle from the CMB data \cite{Minami:2020odp}. It is also well-known that photons travelling significantly large distances, and interacting with magnetic fields, can lead to the production of ALPs (see, for instance, \cite{DeAngelis:2008sk}). Conversions between photons and ALPs, in the presence of  primordial magnetic fields, can also leave observable signatures in the CMB \cite{Mirizzi:2009nq}, which together with other cosmological considerations, has been used to constrain a considerable region of the parameter space \cite{Irastorza:2018dyq}. In the future, we plan to combine the estimate coming from polarization data, and the requirement that ALPs from the early-universe do not decohere, to find new probes for the so-called cosmological axion background \cite{Dror:2021nyr}.

\section{Decoherence through the cosmological medium}

In this section we will look at the potential sources of decoherence of a photon in some quantum state due to the interaction with other particles in the cosmological medium. Unlike for the estimates in the previous section, we know from observations the number density of the other species, with values that make interactions more likely. We already had a first glance at such interactions, like photon-photon scattering with CMB radiation. 

\subsection{Abundance of particles}\label{abpar}

First, we shall compute the number density of photons. This is given by
\begin{align}
& n_{\gamma}=\frac{8 \pi}{c^{3}} \int_{0}^{\infty}\left(\frac{k T}{h}\right)^{3} \frac{x^{2} d x}{e^{x}-1} \nonumber \\
& \implies n_\gamma = 4.11 \times 10^8 \, (1+z)^3  
\, m^{-3}\,,
\end{align}
where the temperature of the CMB is $T_0 = 2.72548\pm 0.00057$K. Other sources give far fewer photons. 

Next, we look at the abundance of baryons. The baryon-to-photon density is 
\begin{equation}
\eta = \frac{n_{\rm b}}{n_\gamma} = 2.75 \times 10^{-8} \Omega_b h^2\,.
\end{equation}
With Planck's (2018) value of $\Omega_{\rm b} h^2 = 0.02237 \pm 0.00015$ \cite{Planck:2018vyg},  this gives an average baryon density today (if fully ionized) of
\begin{equation}
n_{\rm b,0} = 0.2526 \, m^{-3}.
\end{equation}
Primordial nucleosynthesis and the CMB tell us that the Helium-4 mass fraction is about $Y_{\rm P} = 0.246$.  To a good approximation, all the mass is in protons and Helium --- everything else is negligible in terms of number density.

The number density of Helium is given by $Y_{\rm P} = 4n_{\rm He}/(4n_{\rm He}+n_{\rm p})$.   With $Y_{\rm P}=0.246$, $n_{\rm p}/n_{\rm He} = 12.26$.  This means that the fraction of baryonic nuclei that is Helium-4 is 0.0754.   We also have $n_{\rm b} = n_{\rm p} + 4 n_{\rm He} = n_{\rm p}(1+4/12.26) = 1.33 n_{\rm p}$.

Next, the abundance of protons is related to that of baryons by
\begin{equation}
n_{\rm p,0} = \frac{n_{\rm b}}{1.33}  \implies  n_{\rm p} = 0.190 \,(1+z)^3\, m^{-3}\,, z<z_{\rm reion}\,,
\label{np}
\end{equation}
where $z_{\rm reion} \simeq 7.7\pm 0.8$. Apart from protons, essentially all other baryons are Helium-4, which have a number density, after reionization, of
\begin{equation}
n_{\rm He,0} = (0.2526-0.190)/4  \implies  n_{\rm  He} = 0.016\,(1+z)^3 \,m^{-3}\,.
\label{nHe}
\end{equation}
Before reionization (and after recombination) the ionized fraction is about $10^{-4}$, so the proton number density is smaller than equation (\ref{np}) by this factor.  Helium reionization is thought to occur at $z_{\rm He, reion} \simeq 3-4$, although the details remain uncertain. 

The number density of electrons is related to that of protons and Helium-4. Indeed, there is one electron per proton and $2$ per Helium-4, which gives
\begin{equation}
n_{\rm e,0} = 0.190+2 \times 0.016  \implies  n_{\rm e} = 0.222 \,(1+z)^3\,m^{-3}\,,
\end{equation}
after $z_{\rm He, reion}$.  For $z_{\rm He,reion} < z < z_{\rm reion}$, the electron number density would be  $n_{\rm e} = n_{\rm p}$.

\subsection{Interaction with CMB radiation}

Using the cross section for photon-photon scattering, the optical depth is computed as follows
\begin{align}
	{\cal T} & = \frac{1}{2}\int_{t}^{t_0} dt \frac{937\alpha^4 E_{\gamma}^3}{10125\pi m^8} \int_{0}^{\infty} \frac{dE}{2\pi^2} \frac{E^5}{\exp(E/T) - 1} \nonumber \\
	& = \frac{1}{2}\int_{t}^{t_0} dt \frac{937\alpha^4 E_{\gamma}^3}{10125\pi m^8} \frac{1}{2\pi^2} \left(\frac{8\pi^6}{63} T^6 \right) \nonumber \\
	& =  \frac{1874 \pi^3 \alpha^4}{637875} \frac{E_{\gamma,0}^3}{m^8} \int_0^{z} \frac{dz'}{(1+z') H(z')} T_{0}^6 (1+z')^9\,,
\end{align}
where ${\cal T} = 1$ determines the mean free path/redshift of the test photon. The solutions to the equation depend strongly on $E_{\gamma,0}$. For instance, for $E_{\gamma,0} = 10^{-24}\ \Mp \sim 1\ {\rm keV}$, we get{\footnote{We have used $\Omega_m h^2 = 0.1424$, $\Omega_{\Lambda} = 0.6889$ and $\Omega_r h^2 = 4.2\times10^{-5}$.}
\begin{align*}
	\int_0^z dz' \frac{(1+z')^8}{\left[\Omega_{m} (1+z)^3 + \Omega_r (1+z)^4 + \Omega_{\Lambda}\right]^{1/2}} \approx 1.13 \times 10^{34}\,,
\end{align*}
which renders $z \approx 50000$, whereas for $E_{\gamma,0} = 10^{-21}\ \Mp \sim 1\ {\rm MeV}$,
\begin{equation*}
	\int_0^z dz' \frac{(1+z')^8}{\left[\Omega_{m} (1+z)^3 + \Omega_r (1+z)^4 + \Omega_{\Lambda}\right]^{1/2}} \approx 1.13 \times 10^{25}\,,
\end{equation*}
which yields $z \approx 2700$. The conclusion is clear, that the test photon can propagate without interacting with CMB radiation for longer than the latter has been around ($z \sim 1000$). Conversely, let us fix $z = z_{\rm dec} = 1000$, which allows us to compute the probability of the photon travelling without interacting ($P = e^{- {\cal T}(z_{\rm dec})}$). For $E_{\gamma,0} \sim 1\ {\rm keV}$ the probability is essentially $1$, whereas for $E_{\gamma,0}  \sim 1\ {\rm MeV}$, we get $0.9994$, signalling that most likely the test photon would not have interacted from the decoupling era until today due to interaction with CMB radiation. Evidently, one can get less trivial values for higher photon energies. However, notice that at such energies other processes are predominant, which we will not consider for our purposes. 

\subsection{Interactions through Thomson scattering}

Following the same philosophy as before, we will compute the mean free path of a photons interacting through Thomson scattering. The cross section for this process is given by 
\begin{equation}
	\sigma_{\rm th} = \frac{8\pi}{3} \frac{\alpha^2}{m^2}\,,
\end{equation}
where $m$ is the mass of the charged particle. Due to the dependence on this parameter, the interaction with electrons are predominant in comparison with interactions with protons. 
Then, the optical depth is given by
\begin{equation}
	{\cal T} = \frac{8\pi}{3} \frac{\alpha^2}{m^2} \int_0^z \frac{dz'}{(1+z') H(z')} n_e (z')\,,
\end{equation}
where the number density of free electrons is given by
\begin{align}
n_{e,1}(z) &= 0.222 (1+z)^3\ m^{-3}, \quad z<z_{\rm He,reion} \nonumber \\
n_{e,2}(z) &= 0.19 (1+z)^3\ m^{-3}, \quad z_{\rm He,reion} < z < z_{\rm reion}
\end{align}
as computed in the sub-section \ref{abpar}. We are assuming that the ionization fraction is $1$ after reionization and $0$ before it (but after decoupling). With these considerations, the optical depth is
\begin{align}
	{\cal T}  = \frac{8\pi}{3} \frac{\alpha^2}{m^2} & \bigg\{ \int_0^{z_{\rm He,reion}} \frac{dz'\ n_{e,1}(z')}{(1+z') H(z')} \nonumber \\
	& + \int_{z_{\rm He,reion}}^{z_{\rm reion}} \frac{dz'\ n_{e,2}(z')}{(1+z') H(z')} \bigg\}\,.
\end{align}
For the sake of concreteness, we take $z_{\rm He,reion} = 3.5$ and $z_{\rm reion} = 7.8$, yielding
\begin{equation}
	{\cal T} = 0.0529585 \implies P = \exp(-{\cal T}) = 0.94819\,,
\end{equation}
i.e., there is roughly a $95\%$ probability of photons travelling freely from the reionization epoch until present time. Notice that this analysis is basically the same as that for the optical depth for CMB radiation due to the same process. The value obtained by the Planck mission is ${\cal T} = 0.0561 \pm 0.0071$ \cite{Planck:2018vyg}, in good agreement with the result estimated here. Notice that for higher energies one would have to use the Klein-Nishina formula for the cross section; however, it is always less or equal than the Thomson cross section, which renders larger mean free paths. 

\subsection{Other processes}

There are other processes involving photons which could lead to decoherence (or the annihilation of the photon). One such process is the pion photoproduction
\begin{equation}
p + \gamma \longrightarrow	\left( \begin{array}{l} 
	p\\
	n
	\end{array} \right) + \pi\,.
\end{equation}
However, the threshold energies for this kind of processes are very large for our purposes (see \cite{Ruffini:2015oha} for an in-depth study of these processes). Indeed, one should take a closer look to them for energies of order $\sim 10^{15}$ eV or higher. 

On the other hand, as shown in \cite{Berera2020}, x-rays would be more interesting for quantum communication purposes at present day (or low redshifts in general). For said range and due to the dominant constituents of the interstellar medium (photons, electrons and protons) and the weakness of QED, the interactions between photons and the background are negligible. Case in point, the mean free path for interactions with electrons in the interstellar medium was found to be of order $1$ Mpc, which is larger than the size of the Milky Way. Looking at dense regions of the HII gas, the mean free path reduces to $0.1$ kpc, which is a considerable distance within the galaxy. A more in depth discussion of these interactions and others, like with dust particles, galactic magnetic fields, etc., can be found in \cite{Berera2020} and references within. The upshot is that for the (soft) x-ray region of the spectrum, the interactions of a test photon at such energies are rather negligible, as opposed of radio signals which can be affected by galactic magnetic fields, or UHE photons, where particle production and other processes dominate. These conclusions can be extrapolated to low redshifts, as demonstrated in \cite{Mack:2008nv}. That work reports a considerable transparency window for photons in the same energy range ($E \lesssim 10$ keV) for redshifts up to $z \sim 100$. 

\section{Discussion}

In this work we have looked at the possibility of photons maintaining their quantum state over cosmological distances. Naturally, this is an intriguing question in many respects, including quantum communication and quantum teleportation, even more so considering the success achieved in Earth-based experiments. An analysis of quantum coherence to interstellar distances was presented in \cite{Berera2020}, where it was shown that photons in the x-ray range are the prime candidates for these purposes. This work reinforces that conclusion, only now extending to cosmological distances, by generalizing to include other potential sources of decoherence. 

We have used the standard definition that decoherence takes place due to the interaction between a quantum state and an environment. Thus, for expansion-induced decoherence, the question we had to tackle was how gravity can produce an environment. For scalar fields it is well known that squeezing can do the job, where a large number of particles at super-horizon momenta scales are produced. That is not the case for EM fields, because the EM field is conformally invariant and the FLRW metric is conformally flat. The same argument can be applied for the free Dirac theory of fermions. In consequence, there are no excitations of the field owing to gravitational effects and thus no `extra' environment wrecking the quantum state. One obvious loophole consists in breaking the conformal invariance through a coupling with other fields. We have considered ALP-photon interactions, which have been widely studied in the literature. One of the options is a direct interaction which leads to the production of fermions, and the other is the decay of the ALP into two photons, which in turn can interact with the one in a coherent state. In both scenarios the clear conclusion is that interaction rates range from negligibly small to zero, depending on the parameters of the interaction. In some sense we played against our odds by taking large masses for ALPs so that they can decay by the matter-dominated era, or by taking abnormally large energies for the test photon. Regardless of whatever be the case, the conclusion remains the same. We should emphasize that the number densities for ALPs considered are only those produced by expansion, not some primordial population which may lead to an enhanced effect. In fact, axion production by other (standard) mechanisms can lead to a wider variety of more important processes, like the inverse-Primakoff effect, which future radio telescopes could exploit for ALP DM, although this is not expected for the QCD axion \cite{Irastorza:2018dyq}. In consequence, as far as expansion-induced decoherence goes, the results from \cite{Berera2020} can be confidently extrapolated to cosmological distances. 

In order to search for stronger decoherence factors, one has to look at the population of different species in the cosmological medium. Interactions with the CMB radiation is one clear option, where for present-day energies of $~ 1$ keV the interaction rate is essentially 0, whereas for $1$ MeV, the probability of interactions is less than $0.1\%$. Higher probabilities of interaction are associated to Thomson scattering after reionization, where there is roughly a $5\%$ probability of interaction. Thus, for photon energies in the keV range the main decoherence factors lie at galactic scales, where they can maintain the state for considerable distances. Moreover, for the same sweet spot the conclusion holds for low redshifts, or even for $z \sim 100$ \cite{Mack:2008nv}.

In conclusion, the analysis in this paper has examined the free streaming requirements for photons, beyond just the classical condition that they maintain their initial momentum, to the stronger condition that the quantum coherence of the photons is also preserved. Notice for instance that groups of photons initially could also have some form of quantum coherence amongst them through their momentum or internal states, producing for example coherent or lasing states. So, even if the individual momentum of the photons was preserved it is still possible the more delicate quantum coherence amongst the photons could be destroyed. Here we have identified frequency regions in which photon quantum coherence can be maintained up to cosmological scales due to lack of interactions, extending on the work in \cite{Berera2020} that only examined the galactic scale. Recently, the effect of a (curved) Schwarzschild background on the quantum state of coherent light, which can be verified by Earth-to-satellite signals, has been examined \cite{Bruschi:2013sua, Bruschi:2021all, Exirifard:2020yuu}. Building on our present work, it will be natural to consider the effect of accelerating backgrounds on similar coherent wavepackets over cosmological scales and their consequences for quantum communication. 

A natural future direction to follow would be to check the conditions required for the axions to maintain a similar coherent state over cosmological distances. If one can find a similar result that axions also do not decohere due to the background expansion, this opens a new possibility for a complementary signal for axions coming from the very early universe. 

\section*{Acknowledgements}
AB is partially funded by STFC. SB is supported in part by the NSERC (funding reference CITA \#490888-16) through a CITA National Fellowship and by a McGill Space Institute fellowship. RB is supported in part by funds from NSERC and from the Canada Research Chair program and is also grateful to the Institute for Theoretical Physics and the Institute for Particle Physics and Astrophysics of the ETH Zurich for hospitality. JCF is supported by the Secretary of Higher Education, Science, Technology and Innovation of Ecuador (SENESCYT).

\bibliography{apssamp}

\end{document}